# Photonic bandgap fiber bundle spectrometer

**Qu Hang, Bora Ung, Imran Syed, Ning Guo, Maksim Skorobogatiy**

*Génie physique, Ecole Polytechnique de Montreal, C.P. 6079, succ. Centre-ville, Montréal, QC, Canada, H3C 3A7*

*www.photonics.phys.polymtl.ca*

Abstract: we experimentally demonstrate an all-fiber spectrometer consisting of a photonic bandgap fiber bundle and a black and white CCD camera. Photonic crystal fibers used in this work are the large solid core all-plastic Bragg fibers designed for operation in the visible spectral range and featuring bandgaps of 60nm - 180nm-wide. 100 Bragg fibers were chosen to have complimentary and partially overlapping bandgaps covering a 400nm-840nm spectral range. The fiber bundle used in our work is equivalent in its function to a set of 100 optical filters densely packed in the area of ~1cm$^2$. Black and white CCD camera is then used to capture spectrally "binned" image of the incoming light at the output facet of a fiber bundle. To reconstruct the test spectrum from a single CCD image we developed an algorithm based on pseudo-inversion of the spectrometer transmission matrix. We then study resolution limit of this spectroscopic system by testing its performance using spectrally narrow test peaks (FWHM 5nm-25nm) centered at various positions within the 450nm-700nm spectral interval. We find that the peak center wavelength can always be reconstructed within several percent of its true value regardless of the peak width or position. Moreover, we demonstrate that although the widths of the individual Bragg fiber bandgaps are quite large (>60nm) the spectroscopic system has a resolution limit of ~30nm. We conclude by showing theoretically, that, in principle, presented spectroscopic system can resolve much narrower peaks down to several nm in width; this would, however, require a stringent control of the experimental errors during measurement and calibration. We believe that photonic bandgap fiber bundle-based spectrometers have a potential to become an important technology in multispectral imaging because of their simplicity (lack of moving parts), instantaneous response, and high degree of integration.

Key words: Photonic crystal fibers; spectrometer and spectroscopic instrument; fiber optics imaging;

## 1. Introduction

Development of the multispectral imaging systems presents an active research field in rapid development. Inherent advantages of such systems are in the real-time monitoring, cost-effectiveness and high accuracy. Imaging spectrometers have been demonstrated in many applications including chemical component analysis [1], remote sensing [1,2], astronomy [3], and biochemistry [4]. Moreover, multispectral imaging systems have been successfully demonstrated both in the visible and the near inferred spectral ranges [5-9].

A typical multispectral imaging system includes a high sensitivity, high signal-to-noise ratio photo-electric conversion device, such as a CCD array; optical filters or dispersive elements which are either used to provide the reference spectra or to filter-out a specific spectral component; as well as a broadband illuminant (light source). In a typical imaging spectrometer, spectral information is acquired indirectly by taking several images in the complimentary, while possibly overlapping, spectral ranges. A set of intensity images is then converted into spectra by utilizing various interpretation algorithms such as pseudo-inverse estimation (PSE), principal-component analysis (PCA), and nonlinear fitting [5,6].

Recently, several optical fiber-based imaging spectrometers have been reported [7,8]. In [7], for example, the authors used at the input end a 10×10 bundle of the mid infrared fibers which were then separated into a linear 1×100 array at the output end. The output beams were then simultaneously dispersed using a single Bragg grating and imaged on a CCD array. This system, thus, functions as a 10×10 spectral imaging array. In another implementation [8], a multiple range spectrophotometer was implemented using an optical fiber probe. In that system, linear displacement of the light source was projected onto a fiber array (bundle) so that only a single fiber was lit for a given position of the light source. The fiber bundle was then interrogated using a spectrometer, thus enabling spatially resolved real-time data acquisition. Note that in both examples, optical fibers were used exclusively as broadband optical guides, while a separate spectrometer was still required for the spectra acquisition.

The goal of this paper is to demonstrate that the function of a spectrometer can be implemented directly inside a single fiber bundle. Then a complete imaging spectrometer can be realised using an array of such fiber bundles interrogated by a linear or a 2D CCD sensor. Such an approach would allow getting rid of costly and slow traditional spectrometers based on moving gratings, thus leading to significant cost savings and increased acquisition speed. In this paper we report for the first time a photonic bandgap fiber-based spectrometer that consists of a photonic cystal fiber bundle and a black and white CCD camera.

## 2. Characteristics of the subcomponents: Bragg fibers, fiber bundle, and a CCD camera

### *2.1 Photonic bandgap Bragg fibers*

Photonic bandgap fiber is a key element of our spectroscopic system. The photonic bandgap Bragg fibers used in our research are home-made and have been reported previously in [10-12]. An individual Bragg fiber features a large 300um-700um diameter core made of a PMMA plastic. The core region is surrounded with a periodic multilayer reflector featuring ~100 submicron-thick layers of low and high refractive index PMMA/Polystyrene plastics with refractive indices 1.49/1.59. Such a multilayer (Bragg reflector) is responsible for the appearance of the spectrally narrow transmission band (reflector bandgap) within which the light is strongly confined inside the fiber core. For the wavelengths outside of the reflector bandgap the light penetrates deeply into the multilayer region exhibiting strong propagation loss due to scattering on the imperfections inside the multilayer region. A typical fiber loss within the reflector bandgap region is ~10dB/m and is mostly determined by the bulk absorption loss of a low-purity PMMA plastic. Outside of a bandgap region scattering loss dominates resulting in >60dB/m propagation loss. In our experiments we used 30cm-long fibers so that the loss of guided light was below 3dB, while the loss of non-guided light was >20dB. Numerical aperture of all the Bragg fibers was in the range of 0.17-0.22.

To construct a fiber bundle spectrometer we chose 100 Bragg fibers with complimentary and partially overlapping bandgaps as shown in Fig. 1(a). All the fibers in a bundle were drawn from the same preform with the only difference between them being the final diameter. The smaller diameter fibers feature bangaps shifted towards the blue part of a spectrum. In Fig. 1(b) we also show distribution of the fiber bandgap width as a function of the fiber bandgap central wavelength. We note that the fiber bandgap size increases with its center frequency. Thus, in the blue part of a spectrum the fiber bandgap width is ~60nm, while increasing to ~180nm in the red part of a spectrum.

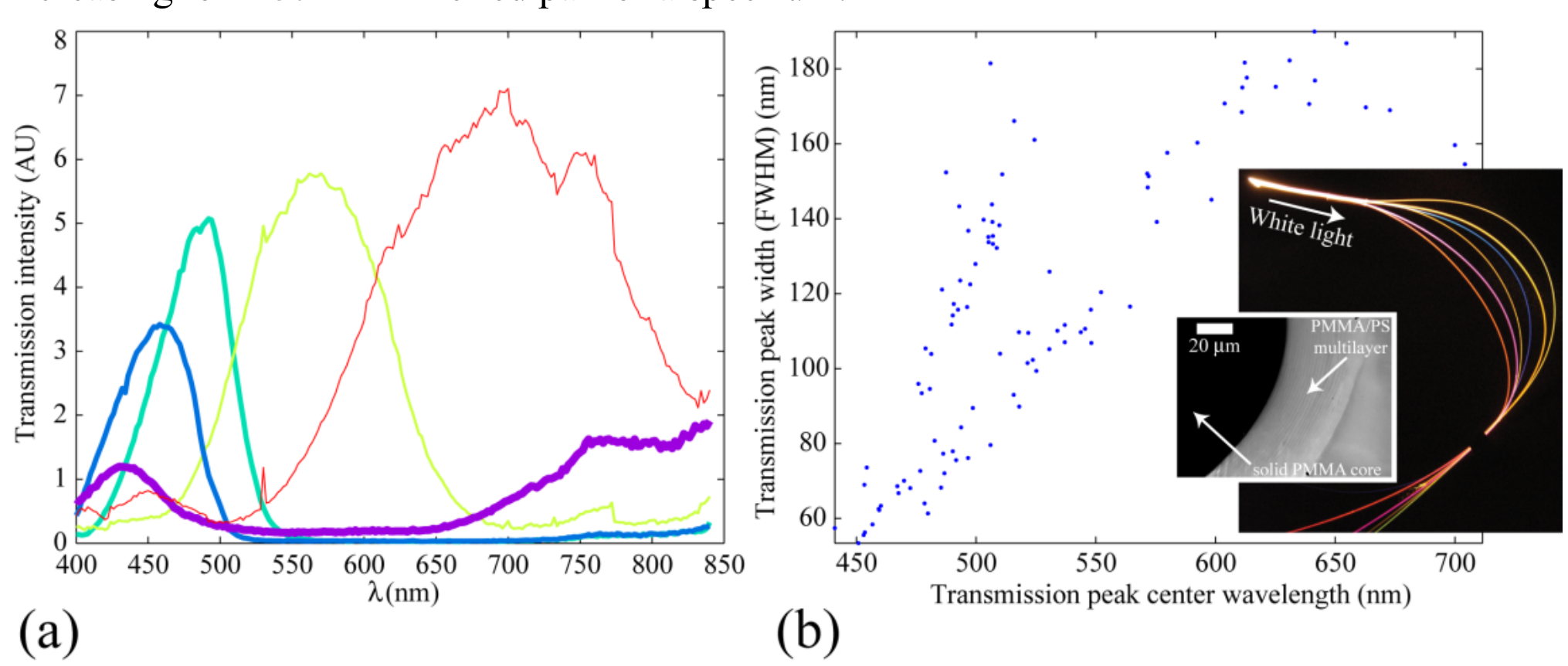

Figure 1. (a) Transmission spectra of 5 typical Bragg fibers used in the fiber bundle. (b) Distribution of the fiber bandgap width as a function of the fiber bandgap center wavelength for all the Bragg fibers in a bundle. In the inset: photo of a Bragg fiber crossection showing a solid core surrounded by a periodic multilayer reflector. When launching white light into the Bragg fiber the non-guided colors are strongly irradiated with only a single color reaching the fiber end.

*2.2 Photonic bandgap fiber bundle*

The fiber bundle used in our experiments featured a 5.6mm inner diameter plastic tube that hosted 100 Bragg fibers of 30 cm of length. Fibers at the input end of a bundle were rigidly attached with epoxy to each other and to the confining tube, and the whole assembly was then polished using optical films of various granularity. On the other end of a bundle 100 fibers were inserted into a custom-made block featuring holes of diameter 0.8mm placed in a periodic 10×10 square array, see Fig. 2. The block, the tube, and all the fibers were bonded together with epoxy, and the output end was then polished.

Principle of operation of a photonic bandgap fiber bundle spectrometer can be clearly understood from Fig. 1 and Fig. 2. In principle, if all the fiber bandgaps were to be spectrally narrow and non-overlapping, then at the end of a fiber bundle the relative intensities of light coming out of the individual fibers would be unambiguously related to the corresponding spectral components of an incoming light. In practice, individual fiber bandgaps are always overlapping, and in our particular implementation the bandgaps are quite broad. Therefore, to reconstruct the intensity of an incoming light from the intensities of light coming out of the individual fibers of a bundle we have to use a certain deconvolution algorithm. Note that spectrometer resolution is not directly limited by the width of fiber bandgaps. As we will demonstrate later in the paper, even with all the individual bandgaps as wide as 60nm one could, in principle, reconstruct the peaks as narrow as 5-10nm. However, to achieve such a fine resolution, management of the experimental becomes critical.

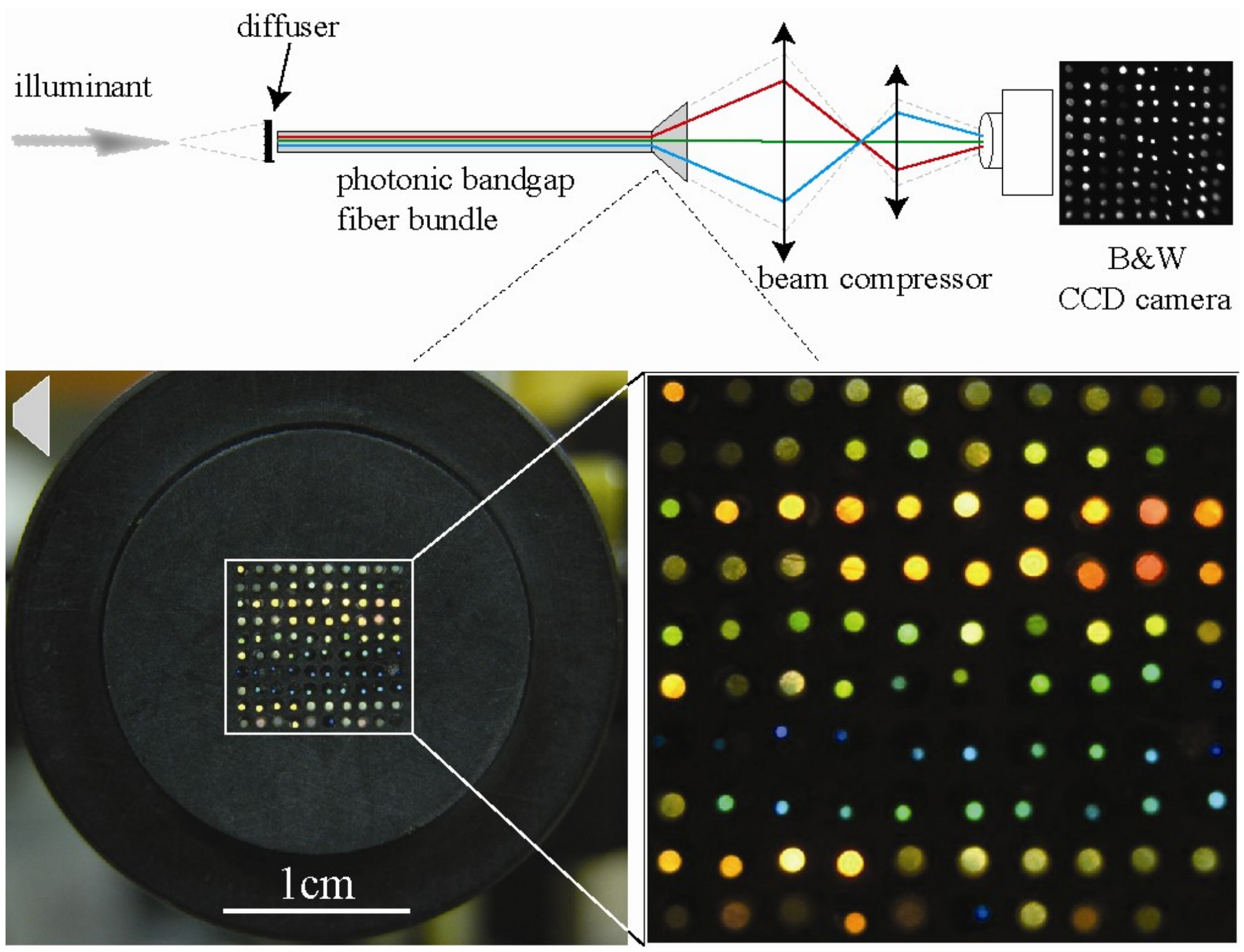

Figure 2. Fiber bundle spectrometer. Top part: schematic of the spectrometer. Light from the illuminant is launched into the fiber bundle. Lower part: when the broadband light is launched into the fiber bundle, the output is a mosaic of colors selected by the individual Bragg fibers. The image is then recorded by the black and white CCD array.

*2.3 Sensitivity and linear response of a CCD sensor.*

The output of the fiber bundle is a 10×10 matrix of coloured fibers which is then projected with a beam compressor onto a black and white CCD array (see Fig. 2). The sensitivity of a CCD sensor would directly affect the measuring range, the signal-to-noise ratio, as well as the algorithm that we use to reconstruct the spectrum of an illuminant. The CCD sensor in our experiment is an Opteon Depict 1, E and S Series B1A 652×494 Black & White CCD. The normalized sensitivity for different wavelengths is displayed in Fig. 3(a) from which it is clear that the sensor covers well all the visible and a part of a near-IR spectral range. Sensitivity curve, however, is a highly variable function of wavelength, which should be taken into account in the re-construction algorithm. Fig. 3(a) was obtained by launching the light form a supercontinuum (SC) source into a monochromator whose output would produce a 2nm-wide (FWHM) peak at a desired center wavelength. Light from such a tuneable source would then be launched into a commercial multimode fiber of ~1mm diameter. The output power from such a fiber would then be measured by both the CCD array and a calibrated power meter; the sensitivity curve is then achieved by dividing one by the other.

As we will see later in the presentation of our paper, one of the key requirements to a CCD array is linearity of its monochromatic response as a function of the intensity of incident light. The near-linear response of a CCD sensor was confirmed at several wavelengths of interest; a typical CCD response (at λ=560nm) is presented in Fig. 3(b). To obtain Fig. 3(b), two polarisers were placed between a CCD array and the output of a commercial fiber in a setup described in the previous paragraph. By varying the angle between the two polarizes we also varied the intensity of light coming onto a CCD array. A reference measurement is performed with a calibrated power meter for the same angles between polarisers; the results are compared to establish a near-linear sensor response.

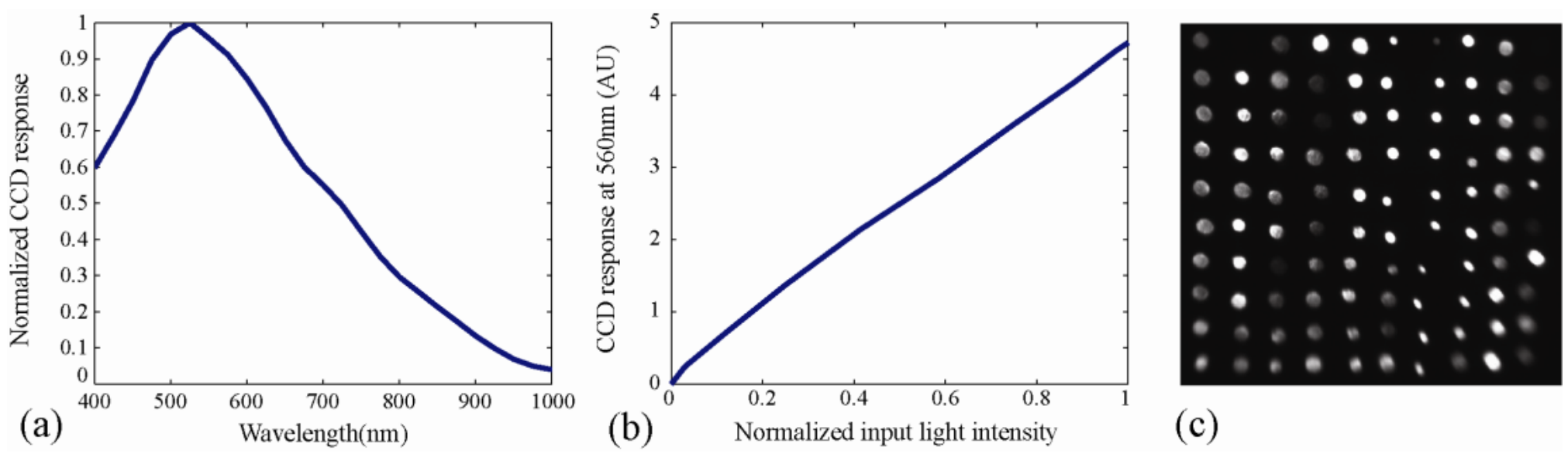

Figure 3. (a). Normalized spectral response of a CCD array. (b) A typical monochromatic near-linear response of a CCD array (λ=560 nm). (c) An image of fiber bundle taken by a CCD array.

**3. Calibration of the fiber bundle spectrometer and spectrum reconstruction algorithm**

In its operation mode the fiber bundle spectrometer is illuminated with a test light, which is then spectrally and spatially decomposed by the fiber bundle see (see Fig. 4). The test spectra used in our experiments were either broadband (halogen lamp source), or narrowband (tuneable monochromator source). Image of the fiber bundle output end was then projected onto a Black and White CCD array using a 2:1 beam compressor. Note that a CCD array does not have to be a 2D camera, instead one can use a much economical linear array with the number of detectors matching the number of Bragg fibers in a bundle. Output of each fiber carries information about the intensity of a certain fraction of an illuminant spectrum as filtering by the individual Bragg fibers. It is reasonable to assume then that if the Bragg fibers of a bundle feature complimentary bandgaps that together cover the spectral range of interest, then the spectrum of an illuminant could be reconstructed.

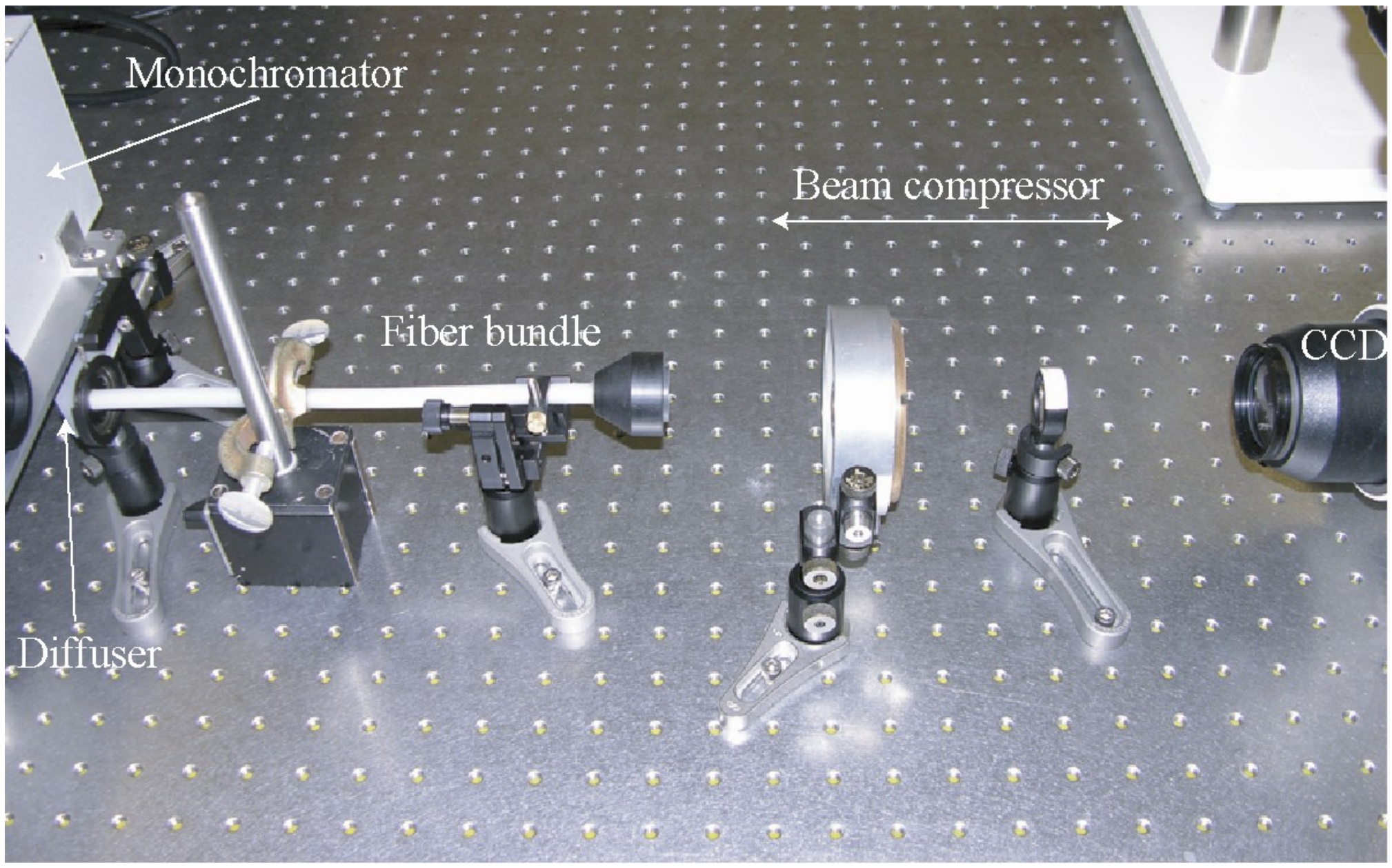

Fig.4. Setup for the spectrometer calibration measurement. Indicated from left to right are the tuneable narrowband monochromator-based source, a diffuser to create a uniform illuminance of the fiber bundle input end, photonic bandgap fiber bundle, a beam compressor made of the two biconvex lens, and a CCD sensor.

*3.1 Transmission matrix method*

To interpret B&W CCD image, we use a so-called luminance adaptation model of the fiber spectrometer. Particularly, for a fixed exposure time of a CCD array, total intensity $C_n$ registered by a CCD sensor from the output of the n$^{th}$ fiber can be presented as:

$$C_n = \int_{\lambda_{\min}}^{\lambda_{\max}} I(\lambda) A_n F_n(\lambda) S(\lambda) O_n(\lambda) d\lambda \quad (1),$$

where $I(\lambda)$ is the illuminant spectral flux density at the fiber-bundle input end, $A_n$ is the effective area of the n$^{th}$ fiber, $F_n(\lambda)$ is the transmission function of the n$^{th}$ fiber, $S(\lambda)$ is the spectral sensitivity of a CCD array, $O_n(\lambda)$ is a fiber-position dependent transmission function of various optics (diffuser, beam compressor, and a CCD objective). Measuring such transmission functions individually is a daunting task. Instead, we use a calibration procedure that measures a compounded transmission matrix of our spectrometer defined as $T_n(\lambda) = A_n F_n(\lambda) S(\lambda) O_n(\lambda)$ . The model equation (1) could then be now rewritten as:

$$C_n = \int_{\lambda_{\min}}^{\lambda_{\max}} I(\lambda) T_n(\lambda) d\lambda = \sum_{i=0,\ \lambda_i=\lambda_{\min}+i\Delta\lambda}^{N=(\lambda_{\max}-\lambda_{\min})/\Delta\lambda} T_n(\lambda_i) \int_{\lambda_i}^{\lambda_i+\Delta\lambda} I(\lambda) d\lambda = \sum_{i=0}^{N} I_i \cdot T_n^i \quad (2).$$

Experimentally, an integral in equation (2) is rather a sum over the intensity contributions from small spectral "bins" of size $\Delta\lambda$. Equation (2) presented in the matrix form becomes:

$$\begin{bmatrix} C_1 \\ C_2 \\ \cdot\cdot \\ \cdot\cdot \\ C_{100} \end{bmatrix} = \begin{bmatrix} T_{1,1} & T_{1,2} & \cdots & T_{1,N} \\ T_{2,1} & T_{2,2} & \cdots & T_{2,N} \\ \cdot\cdot & & & \\ \cdot\cdot & & & \\ T_{100,1} & T_{100,2} & \cdots & T_{100,N} \end{bmatrix} \begin{bmatrix} I_1 \\ I_2 \\ \cdot\cdot \\ \cdot\cdot \\ I_N \end{bmatrix} \quad (3),$$

where vector $(C)_{100\times1}$ represents the intensities of light coming out of the individual fibers as measured by the CCD sensor, $(T)_{100\times N}$ is a spectrometer transmission matrix, and $(I)_{N\times1}$ is a discretized spectrum of the illuminant. From (3) it follows that a discretized spectrum of the illuminant can be reconstructed from the corresponding CCD image ($C$ vector) by inverting the transmission matrix of a spectrometer.

*3.2. Calibration measurement, building a transmission matrix*

To construct transmission matrix experimentally we note that if the illuminant spectrum is monochromatic ($I_i = 0,\ \forall i \neq i_\lambda$ in (3)), then the measured $C^{i_\lambda}$ vector is proportional to the $i_\lambda$ column of the transmission matrix:

$$C^{i_\lambda}(1:100) = T(1:100, i_\lambda) \cdot I_{i_\lambda} \quad (4).$$

To construct the transmission matrix experimentally we use a tuneable monochromator-based narrow-band (2nm FWHM) source to generated "monochromatic" spectra (see Fig. 4). Particularly, we vary the source center wavelength in 2nm increments, thus effectively subdividing the 400nm-840nm spectral interval under consideration into $N = 221$ equivalent 2nm-wide bins. For every new position of the source center wavelength we acquire a $C$ vector using a CCD array and consider it as the next column of the spectrometer transmission matrix. Finally, to finish calibration we measure the wavelength-dependent intensities $I_i$ of a tuneable source by placing a calibrated power meter directly at the output of a monochromator. By dividing every $C$ vector by the corresponding $I_i$ value the transmission matrix is constructed.

In our experiments an S&Y halogen lamp source was used with a Newport Oriel 1/8m monochromator to build a narrow-band tuneable source of 2.0nm FWHM (see Fig. 4). The light beam from a source was then directed onto a diffuser placed right before the fiber bundle to guarantee a uniform illumination of its input end. At the output end of a fiber bundle a 2:1 beam compressor was used to image the fiber bundle output facet onto an 8-bit B&W CCD sensor array. Individual images were then interpreted using a Matlab image processing suite to construct the corresponding $C$ vectors. Wavelength dependent intensity of a tuneable source was measured using a calibrated Newport 841-PE power meter.

*3.3 Spectral reconstruction algorithm*

As it was noted earlier, a discretized spectrum of the illuminant can be reconstructed from the corresponding CCD image ($C$ vector) by inverting the transmission matrix of a spectrometer in (3). However, an immediate problem that one encounters when trying to invert the transmission matrix is that the matrix is non-square. Even if the number of spectral bins is chosen to match the number of fibers in the bundle, thus resulting in a square transmission matrix, one finds that such a matrix is ill-conditioned. One, therefore, has to resort to an approximate inverse of a transmission matrix. To find a pseudo-inverse of a transmission matrix in equation (3) we employ a singular value decomposition (SVD) algorithm. Particularly, form the linear algebra we know that any (100×N) matrix (suppose that N>100) can be presented in the form:

$$\begin{gathered}(T)_{100\times N} = (U)_{100\times 100}(S)_{100\times N}\left(V^T\right)_{N\times N} \\ U^T U = 1\ ;\ V^T V = 1\ ;\ S = diag(\sigma_1, \sigma_2, \sigma_3, \ldots \sigma_{100}) \quad (4), \\ \sigma_1 > \sigma_2 > \sigma_3 > \ldots > \sigma_{100} > 0\end{gathered}$$

where matrixes $U$ and $V$ are unitary, and matrix $S$ is a diagonal matrix of the real positive singular values. For the ill-conditioned matrices, most of the singular values are small and can be taken as zero. By limiting the number of non-zero singular values to $N_\sigma$, equation (4) can be rewritten as:

$$\begin{gathered}(T)_{100\times N} = (U)_{100\times N_\sigma}(S)_{N_\sigma\times N_\sigma}\left(V^T\right)_{N_\sigma\times N} \quad (5), \\ \sigma_1, \sigma_2, \ldots, \sigma_{N_\sigma} \neq 0\end{gathered}$$

and a corresponding pseudo-inverse of the transmission matrix is then found as:

$$(T)^{-1}_{N\times 100} = (V)_{N\times N_\sigma}(S)^{-1}_{N_\sigma\times N_\sigma}\left(U^T\right)_{N_\sigma\times 100} \quad (6).$$

**5. Experimental results**

To test our fiber bundle spectrometer we consider resolving a set of the 25nm-wide peaks centered at 450nm, 500nm, 550nm, 600nm, and 700nm. Such peaks were created using the same monochromator-based tuneable source as used for the spectrometer calibration, however adjusted to have a 25nm bandwidth of the outgoing light. In Fig. 5 we demonstrate in black dashed curves the test spectra of light beams at the input of a fiber bundle (as resolved by another Oriel monocromator), while in thick red lines we show the corresponding reconstructed spectra using the transmission matrix inversion algorithm. In the figures we also indicate the optimal number of non-zero singular values used in the matrix inversion procedure (6). The choice of the number of non-zero singular values used in the spectrum reconstruction algorithm effects strongly the quality of a reconstructed spectrum. Note, in particular, that although the spectral intensity function should be strictly non-negative, the reconstructed spectral intensity can take negative values. Therefore, spectrum reconstruction error can be defined as the ratio of the most negative value of the reconstructed spectral intensity to its most positive value:

$$Error(N_\sigma) = -\min_\lambda(I_{N_\sigma}^{reconstr.}) / \max_\lambda(I_{N_\sigma}^{reconstr.}) \quad (7),$$

and the optimal $N_\sigma$ to be used in (6) is the one that minimises the error (7).

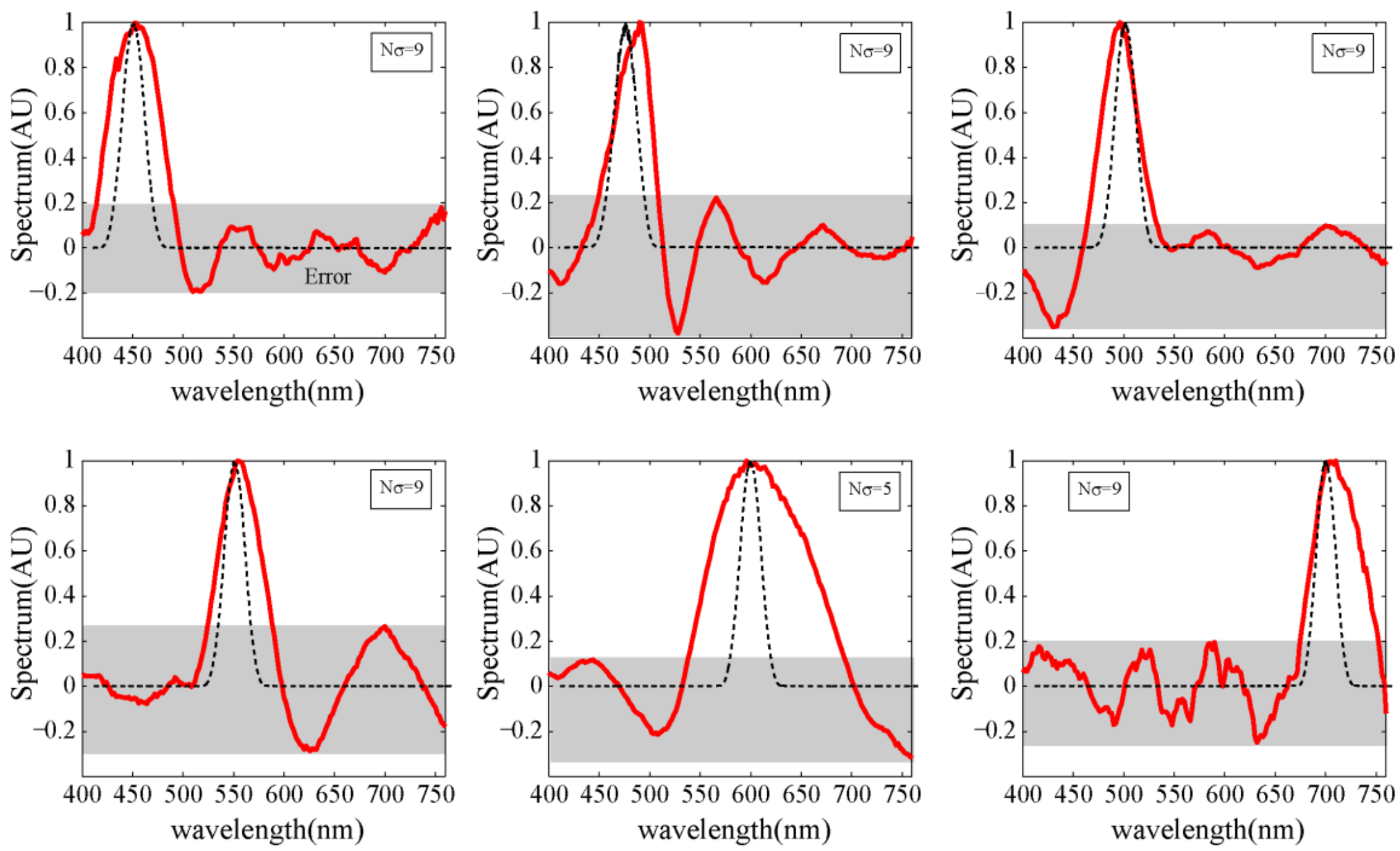

Figure 5. Spectra reconstruction using photonic bandgap fiber bundle-based spectrometer. In black dashed curves are the test spectra of the incoming light as created by the tuneable monochromator-based source. In red solid curves are the corresponding spectra reconstructed by the fiber bundle spectrometer. Nσ indicates the optimal number of singular values used in the transmission matrix inversion algorithm. Gray areas indicate error levels.

From Fig. 5 we conclude that the test peak position (the value of the peak center wavelength) can be reconstructed within several percent of its true value in the whole spectral rage covered by the spectrometer. Moreover, we note that although the widths of the individual Bragg fiber bandgaps are quite large (>60nm) the fiber bundle-based spectroscopic system can resolve peaks of much smaller spectral width. Indeed, in Fig. 5, the widths of the reconstructed spectra are in the 30-50nm range with the exception of a peak centered at 600nm. In the next section we present theoretical estimates for the spectral resolution limit of our fiber bundle spectrometer. We mention in passing that we have also conducted measurements using 5nm, 10nm, 15nm, and 20nm-wide peaks. In all the experiments we saw that the center wavelength of even the narrowest peak can be reconstructed with high precision, however the reconstructed peak width always stayed larger than 30nm.

### 6. Limits of spectral resolution for the fiber bundle-based spectrometer

To understand the resolution limit of our spectrometer we first study the effect of the choice of the number of non-zero singular values $N_\sigma$ used in the inversion algorithm (6). At first we assume that there is no noise in the system. As a test spectrum $I(\lambda)$ we consider a 20nm-wide peak centered at the wavelength of 520nm (dashed black curve in Fig 6(a)).

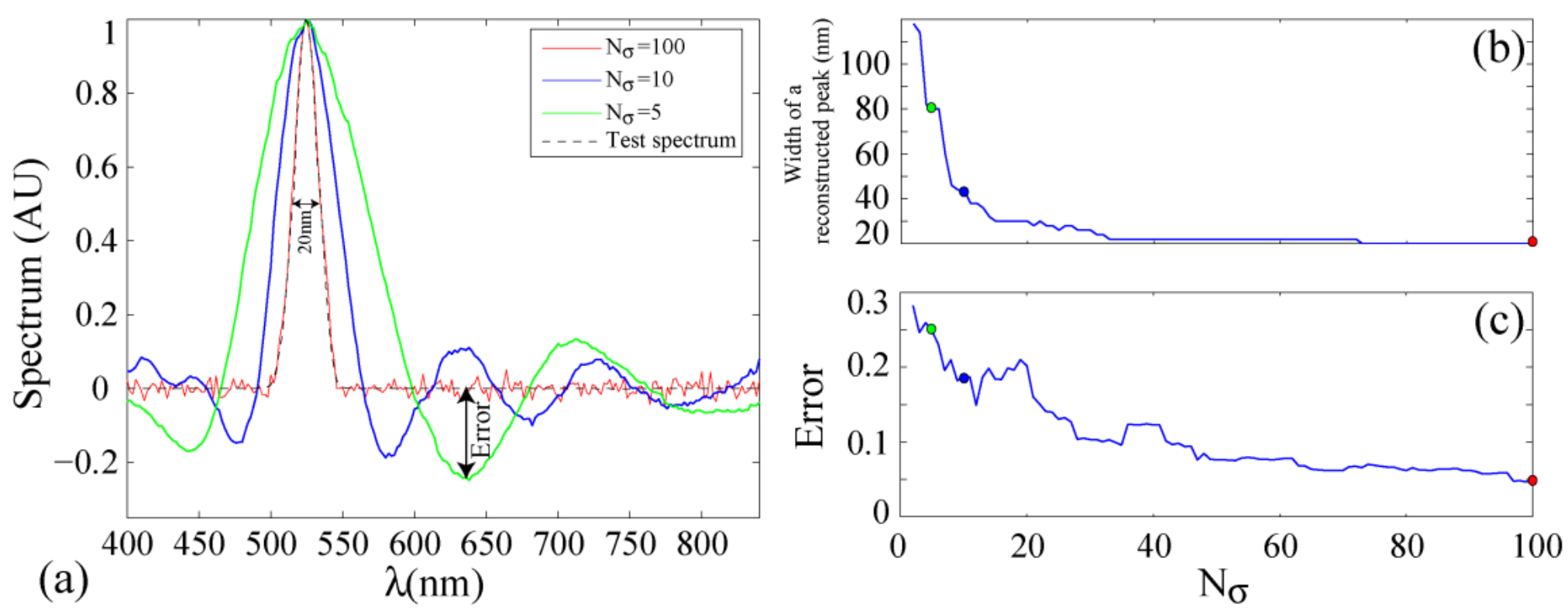

Figure 6. Properties of the reconstructed spectra as a function of the number of singular values used in the inversion algorithm. No noise is present in the system. (a) Dependence of the spectral shape of a reconstructed peak on $N_\sigma$. (b) Width of a reconstructed peak as a function of $N_\sigma$. (c) Reconstruction error as a function of $N_\sigma$.

We then multiply the discritized test spectrum by the transmission matrix of our spectrometer (measured experimentally), and then reconstruct the test spectrum by using a transmission matrix pseudo-inverse with $N_\sigma$ singular values:

$$\begin{gathered}(C)_{100\times1}=(T)_{100\times N}(I)_{N\times1}\\ \left(I_{N_\sigma}^{reconstr.}\right)_{N\times1}=(V)_{N\times N_\sigma}(S)^{-1}_{N_\sigma\times N_\sigma}\left(U^T\right)_{N_\sigma\times100}(C)_{100\times1}\end{gathered}\quad(8).$$

Thus reconstructed spectrum is then compared to the original test spectrum. In Fig. 6(a) we present several reconstructed spectra for the different number of singular values used in the inversion algorithm. If all the 100 singular values are used (red thin curve in Fig. 6(a)), then the 20nm-wide peak is very well reconstructed featuring less than 5% intensity difference with the reconstructed spectrum. The error manifests itself in the form of ripples at the peak tails. When smaller number of singular values are used (for example $N_\sigma = 10$, blue solid cure in Fig. 6(a)), center wavelength of the reconstructed peak still coincides very well with that of a test peak, however, reconstructed peak width is larger than that of a narrow test peak. This result is easy to understand by remembering that expansion basis set used in the reconstruction algorithm is formed by the relatively broad (>60nm see Fig. 1) transmission spectra of the Bragg fibers, therefore one needs a large number of such basis functions (singular values) to reconstruct a narrow spectral feature. In Fig. 6(b) the FWHM of the reconstructed peak is shown as a function of the number of singular values used in the matrix inversion. Note that more than 40 singular values have to be used to achieve the width of a reconstructed peak to be close to 20nm width of a test peak. When only 10 singular values are used, the width of a reconstructed peak increases to 40nm. Additionally, the peak reconstruction error as defined by (7) increases to 20% when a small number of singular values $N_\sigma < 20$ is used. From this we conclude that with no noise present in the system, the peak reconstruction error and the quality of a reconstructed spectrum improves when a larger number of singular values is used.

We now study the effect of experimental noise on the quality of a reconstructed spectrum. There are several sources of noise in our spectrometer. One is the discrete intensity resolution of a CCD array; with the 8-bit encoding the fundamental noise is ~1%. Moreover, because of the relatively large size of a fiber bundle (~6mm) it is difficult to ensure the same illumination conditions of the bundle input facet for all the sources used in the experiments.

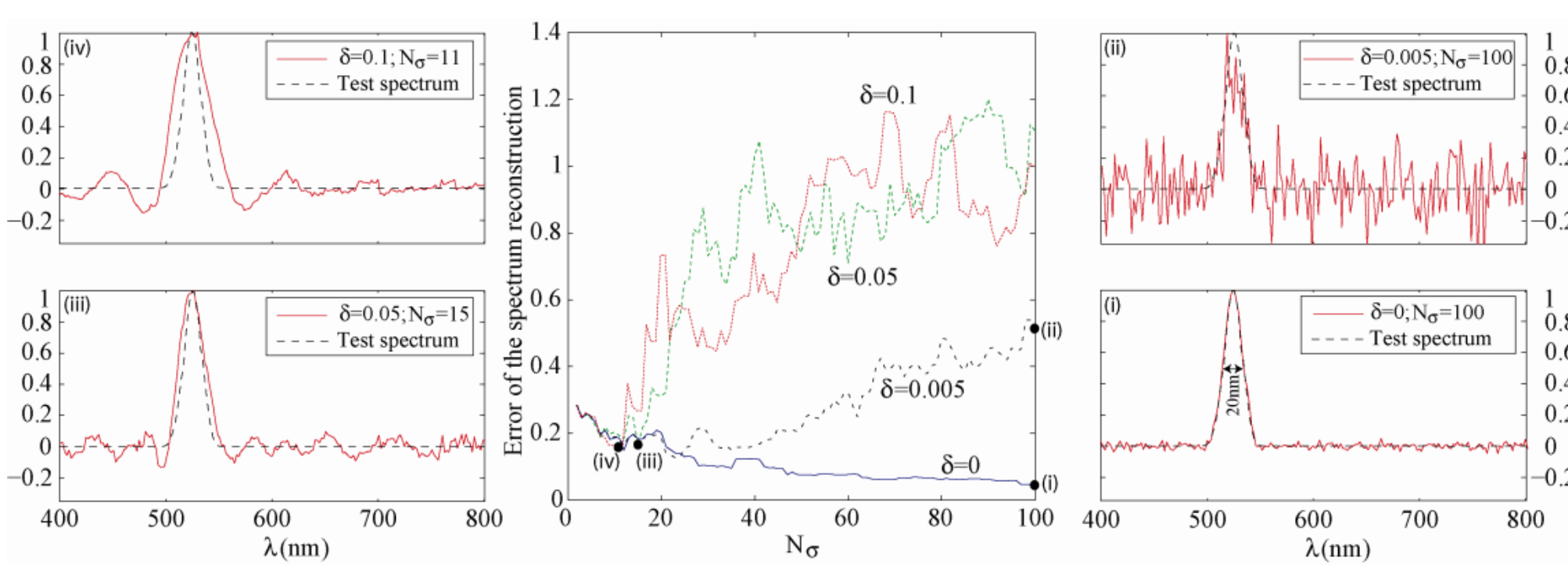

Figure 7. Effect of noise on the quality of reconstruction. Examples of the reconstructed spectra for several particular realisations of noise with amplitudes: (i) $\delta = 0$, (ii) $\delta = 0.005$, (iii) $\delta = 0.05$, (iv) $\delta = 0.1$.

To model experimental noise numerically, we first multiply the test spectrum by the transmission matrix of our spectrometer, then add a uniformly distributed random noise of the relative intensity $\delta$ to the $C$ vector (B&W image), and finally reconstruct the test spectrum by using a transmission matrix pseudo-inverse with $N_\sigma$ singular values:

$$\begin{gathered}(C)_{100\times1} = (T)_{100\times N}(I)_{N\times1} \\ (C^{noise})_{100\times1} = (C)_{100\times1}\cdot\left(1+\delta\cdot(\eta)_{100\times1}\right)\ ;\ \eta-random\subset[-0.5,0.5] \\ \left(I_{N_\sigma}^{reconstr.}\right)_{N\times1} = (V)_{N\times N_\sigma}(S)^{-1}_{N_\sigma\times N_\sigma}\left(U^T\right)_{N_\sigma\times100}\left(C^{noise}\right)_{100\times1}\end{gathered} \quad (9).$$

With these definitions, we first revisit the case when all the 100 singular values are used for spectrum reconstruction. As noted above, in the absence of noise ( $\delta = 0$ in Fig. 7(i)) the 20nm peak can be well reconstructed with only ~5% error. However, addition of even a small amount of noise ( $\delta = 0.005$ in Fig. 7(ii)) results in a highly noisy reconstructed image with errors as large as 50%. Note that by using a smaller number of singular values $N_\sigma \subset [20,40]$ (see the error plot in the middle of Fig. 7) the error of the peak reconstruction can be greatly reduced down to 10%, while still allowing a fair estimate of the peak width ~30nm. When the noise level is increased further, to avoid large reconstruction errors one has to use a relatively small number of singular values $N_\sigma < 20$ (see Figs. 7(iii,iv)). Notably, reconstruction error saturates at 20% even for large noise levels, however, the reconstructed peak width stays always larger than 30nm due to small number of singular values used.

Finally, Fig. 8 presents statistical averages of various parameters (thick solid lines), as well as their statistical deviation from the average (thin dashed curves) as a function of the noise level. To construct these curves, for every value of the noise amplitude $\delta$ we, first, generate 200 realizations of noise. Then, for every realisation of noise we plot the error of peak reconstruction as a function of the number of singular values (similar to Fig. 7). Lastly, the smallest error, the corresponding $N_\sigma$ and the width of a reconstructed peak are recorded and statistical averages are performed. In Fig. 8(a) we present the average peak reconstruction error as a function of the noise amplitude. Note that although the error grows with the noise amplitude, it, nevertheless, saturates at ~20% even for large noise levels. In the inset of Fig. 8(a) we see that at very small noise levels ( $\delta < 0.02$ ) to obtain the smallest reconstruction error one can use almost all 100 singular values, while at higher noise levels less than 20 singular values should be used. As a consequence, for larger noise amplitudes the width of a reconstructed peak could become considerably larger than that of a test peak (20nm).

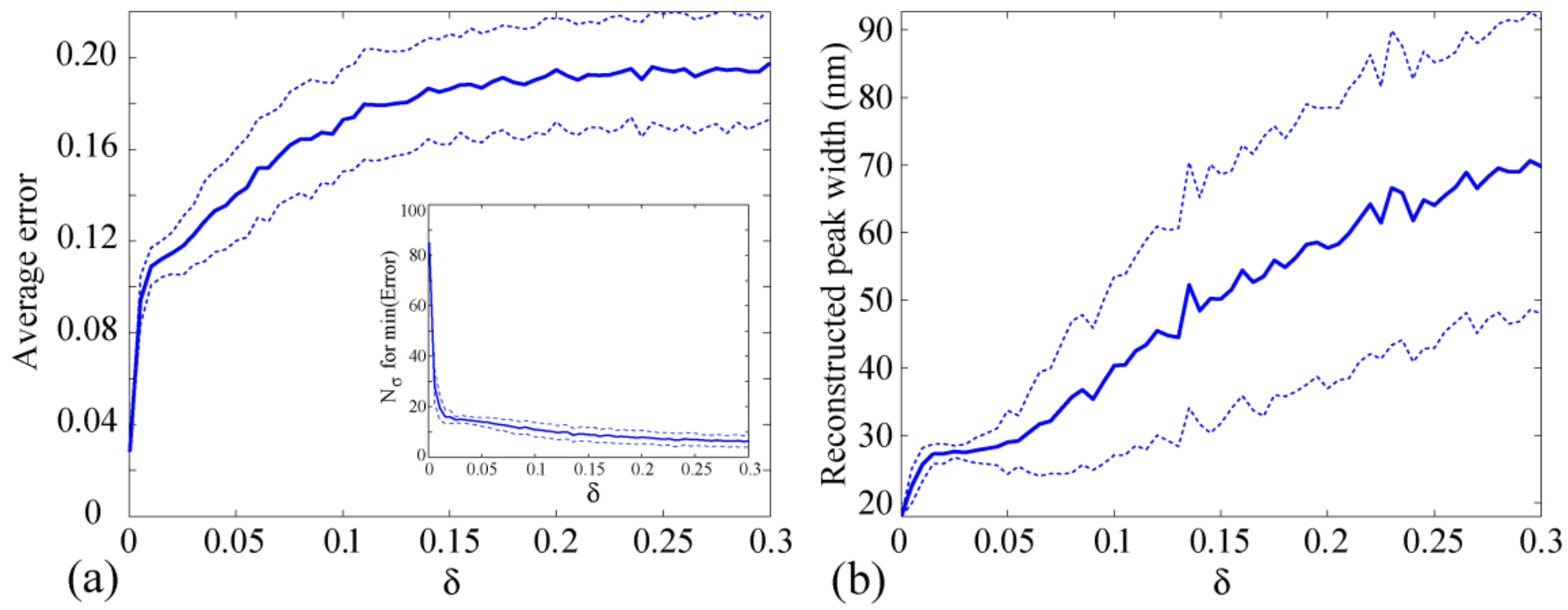

Figure 8. Effect of noise on the reconstruction algorithm. (a) Average reconstruction error and its statistical deviation as a function of the noise amplitude. Inset: optimal number of singular values needed to minimise the reconstruction error. (b) Average width of a reconstructed peak and its statistical deviation as a function of the noise level.

## 7. Conclusion

We have demonstrated experimentally a novel all-fiber spectrometer comprising a photonic bandgap fiber bundle and a black and white CCD camera. 100 Bragg fibers with complimentary bandgaps covering a 400nm-840nm spectral range were placed together to make a ~6mm in diameter, 30cm-long fiber bundle. The spectrometer operates by launching the light into a fiber bundle and then recording the color-separated image at the fiber bundle output facet using a CCD camera. Inversion algorithm was developed to reconstruct the test spectrum using black and white intensity images. Among the clear advantages of our spectrometer is the lack of moving parts, a near instantaneous and parallel acquisition of all the spectral components, compactness, high degree of integration and simplicity of operation.

Photonic bandgap fiber bundle spectrometer was demonstrated experimentally to resolve consistently and with high precision the center wavelengths (positions) of the relatively narrow (5nm-25nm) spectral peaks. If the peak width is much narrower than the bandwidth of Bragg fibers (>60nm) making the fiber bundle, the peak width can still be resolved correctly, however, the reconstruction algorithm is sensitive to experimental noise. Experimental resolution limit of the fiber bundle spectrometer presented in this paper is found to be ~30nm. Theoretical analysis predicts that resolution limit of the existing setup could be greatly improved via careful management of the experimental errors.

## References and links